\def\etc{{\it etc.}}

\def\ie{{\it i.e.}}

\def\~{{$\tilde{\phantom{a}}$}}

\documentclass [12pt] {article}
\usepackage{epsfig}
\usepackage{color}

\def\thebibliography#1{\section{References}\markboth
 {REFERENCES}{REFERENCES}\list
 {[\arabic{enumi}]}{\settowidth\labelwidth{[#1]}\leftmargin\labelwidth
 \advance\leftmargin\labelsep
 \usecounter{enumi}}
 \def\newblock{\hskip .11em plus .33em minus -.07em}
 \sloppy
 \sfcode`\.=1000\relax}
\def\upcite#1{\raise6pt\hbox{\scriptsize
\cite{#1}}}
  \def\lsim{\mathrel {\vcenter {\baselineskip 0pt \kern 0pt
    \hbox{$<$} \kern 0pt \hbox{$\sim$} }}}
    \def\gsim{\mathrel {\vcenter {\baselineskip 0pt \kern 0pt
    \hbox{$>$} \kern 0pt \hbox{$\sim$} }}}

\def\hline{\noalign{\hrule \vskip2pt}}

\def\|{\ifmmode\Vert\else \char`\|\fi}
\ifx\oldzeta\undefined                          
  \let\oldzeta=\zeta                            
  \def\zzeta{{\raise 2pt\hbox{$\oldzeta$}}}     
  \let\zeta=\zzeta                              
\fi

\ifx\oldchi\undefined                           
  \let\oldchi=\chi                              
  \def\cchi{{\raise 2pt\hbox{$\oldchi$}}}       
  \let\chi=\cchi                                
\fi

\def\frac#1#2{{#1 \over #2}}

\def\half{\ifinner {\scriptstyle {1 \over 2}}
   \else {1 \over 2} \fi}

\def\abs#1{\left\vert#1\right\vert}	

\def\simge{\mathrel{%
   \rlap{\raise 0.511ex \hbox{$>$}}{\lower 0.511ex \hbox{$\sim$}}}}
\def\simle{\mathrel{
   \rlap{\raise 0.511ex \hbox{$<$}}{\lower 0.511ex \hbox{$\sim$}}}}

\def\buildchar#1#2#3{{\null\!                   
   \mathop#1\limits^{#2}_{#3}                   
   \!\null}}                                    
\def\overcirc#1{\buildchar{#1}{\circ}{}}

\def\slashchar#1{\setbox0=\hbox{$#1$}           
   \dimen0=\wd0                                 
   \setbox1=\hbox{/} \dimen1=\wd1               
   \ifdim\dimen0>\dimen1                        
      \rlap{\hbox to \dimen0{\hfil/\hfil}}      
      #1                                        
   \else                                        
      \rlap{\hbox to \dimen1{\hfil$#1$\hfil}}   
      /                                         
   \fi}                                         %

\def\subrightarrow#1{
  \setbox0=\hbox{
    $\displaystyle\mathop{}
    \limits_{#1}$}
  \dimen0=\wd0
  \advance \dimen0 by .5em
  \mathrel{
    \mathop{\hbox to \dimen0{\rightarrowfill}}
       \limits_{#1}}}                           

\def\overlay#1#2{\ifmmode%
\setbox0=\hbox{$#1$}%
\setbox1=\hbox to\wd0{\hss$#2$\hss}\else%
\setbox0=\hbox{#1}%
\setbox1=\hbox to\wd0{\hss#2\hss}\fi%
#1\hskip-\wd0\box1 }

\def\pmb#1{\leavevmode\setbox0=\hbox{#1}%
\kern-.02em\copy0\kern-\wd0
\kern.04em\copy0\kern-\wd0
\kern-.02em\raise.04em\box0 }

\def\vereq#1#2{\lower3pt\vbox{\baselineskip1.5pt \lineskip1.5pt
\ialign{$\m@th#1\hfill##\hfil$\crcr#2\crcr\sim\crcr}}}

\def\tensor#1{\protect\@ontopof{#1}{\leftrightarrow}{1.15}\mathord{\box2}}
\def\overstar#1{\protect\@ontopof{#1}{\ast}{1.15}\mathord{\box2}}
\def\overdots#1{\protect\@ontopof{#1}{\cdots}{1.0}\mathord{\box2}}
\def\overcirc#1{\protect\@ontopof{#1}{\circ}{1.2}\mathord{\box2}}
\def\loarrow#1{\protect\@ontopof{#1}{\leftarrow}{1.15}\mathord{\box2}}
\def\roarrow#1{\protect\@ontopof{#1}{\rightarrow}{1.15}\mathord{\box2}}

\def\@ontopof#1#2#3{%
{\mathchoice
{\@@ontopof{#1}{#2}{#3}\displaystyle\scriptstyle}%
{\@@ontopof{#1}{#2}{#3}\textstyle\scriptstyle}%
{\@@ontopof{#1}{#2}{#3}\scriptstyle\scriptscriptstyle}%
{\@@ontopof{#1}{#2}{#3}\scriptscriptstyle\scriptscriptstyle}%
}%
}

\def\@@ontopof#1#2#3#4#5{%
\setbox0=\hbox{$#4#1$}%
\setbox1=\hbox{$#5#2$}%
\setbox2=\hbox{}\ht2=\ht0 \dp2=\dp0 %
\ifdim\wd0>\wd1 %
\setbox1=\hbox to\wd0{\hss\box1\hss}%
\mathord{\rlap{\raise#3\ht0\box1}\box0}%
\else   %
\setbox1=\hbox to.9\wd1{\hss\box1\hss}%
\setbox0=\hbox to\wd1{\hss$#4\relax#1$\hss}%
\mathord{\rlap{\copy0}\raise#3\ht0\box1}%
\fi
}%

\def\lambdabar{\protect\@lambdabar}
\def\@lambdabar{%
\relax
\bgroup
\def\@tempa{\hbox{\raise.73\ht0
\hbox to0pt{\kern.25\wd0\vrule width.5\wd0
height.1pt depth.1pt\hss}\box0}}%
\mathchoice{\setbox0\hbox{$\displaystyle\lambda$}\@tempa}%
{\setbox0\hbox{$\textstyle\lambda$}\@tempa}%
{\setbox0\hbox{$\scriptstyle\lambda$}\@tempa}%
{\setbox0\hbox{$\scriptscriptstyle\lambda$}\@tempa}%
\egroup
}

\def\corresponds{{\lower.2ex\hbox{=}}{\rm\kern-.75em^\triangle}}
\def\succsim{\succ\kern-.9em_\sim\kern.3em}
\def\precsim{\prec\kern-1em_\sim\kern.3em}
\def\slantfrac#1#2{\kern1em^{#1}\kern-.3em/\kern-.1em_{#2}}

\begin{document}
                                                                
\begin{center}
{\Large\bf Axicon Gaussian Laser Beams}
\\

\medskip

Kirk T.~McDonald
\\
{\sl Joseph Henry Laboratories, Princeton University, Princeton, NJ 08544}
\\
(March 14, 2000)
\end{center}

\section{Problem}

Deduce an axicon solution for a Gaussian laser beam in vacuum, \ie, a
beam with radial polarization of the electric field.

\section{Solution}

\begin{figure}[htp]  
\begin{center}
\includegraphics[width=6in, angle=0, clip]{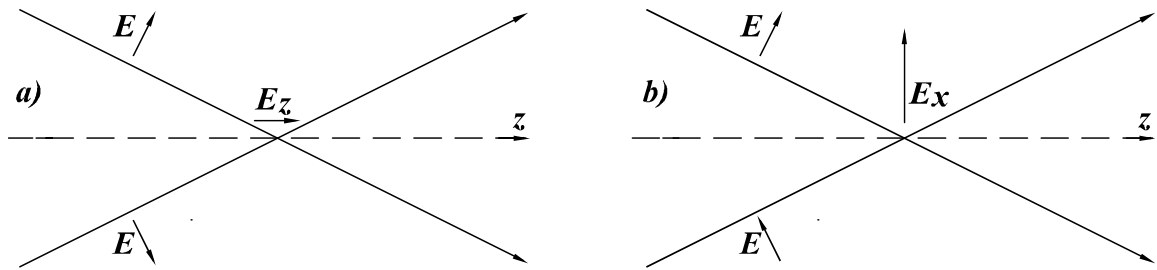}
\parbox{5.5in} 
{\caption[ Short caption for table of contents ]
{\label{axiconfig} a) The radial polarization of the electric field of an 
axicon laser beam leads to a longitudinal electric field at the focus.
b) For a linearly polarized laser beam, shown here with polarization along the
$x$ axis, the electric field is transverse at the 
focus.
}}
\end{center}
\end{figure}

If a laser beam is to have radial transverse polarization, the transverse
electric must vanish on the symmetry axis, which is charge free in vacuum.
However,
we can expect a nonzero longitudinal electric field on the axis, noting 
that the projections onto the axis of the electric field vectors of rays
 all have the same sign, as shown in Fig.~\ref{axiconfig}a.
This contrasts with the case of linearly polarized Gaussian laser beams 
\cite{Siegman,Eberly,Yariv,diff} for which rays at $0^\circ$ and $180^\circ$ 
azimuth to the polarization direction have axial electric field components of
opposite sign, as shown in Fig.~\ref{axiconfig}b. 
The longitudinal electric field of axicon laser beams 
may be able to
transfer net energy to charged particles that propagate along the optical axis,
providing a form of laser acceleration \cite{Pantell,Caspers,Bochove,Kimura}.

Although two of the earliest papers on Gaussian laser beams
\cite{Goubau,Boyd} discuss axicon modes (without using that term, and without
deducing the simplest axicon mode), most 
subsequent literature has emphasized linearly polarized Gaussian beams.
We demonstrate that a calculation that begins with the vector potential 
(sec.~2.1) leads to both
the lowest-order linearly polarized and axicon modes.
We include a discussion of Gaussian laser pulses as well as continuous beams,
and find in sec.~2.2 that the temporal pulse shape must obey condition 
(\ref{e8}).
The paraxial wave equation and its lowest-order, linearly polarized solutions 
are reviewed in secs.~2.3-4.
Readers familiar with the paraxial wave equation for linearly polarized
Gaussian beams may wish to skip directly to sec.~2.5 in which the axicon mode
is displayed.  In sec.~2.6 we find an expression for a guided axicon beam, \ie, 
one that requires a conductor along the optical axis.

\subsection{Solution via the Vector Potential}

Many discussions of Gaussian laser beams emphasize a single electric field
component such as $E_x = f(r,z)e^{i(k z - \omega t)}$ of a cylindrically 
symmetric beam of angular frequency $\omega$ and wave number $k = \omega/c$
propagating in vacuum along the $z$ axis.  
Of course, the electric field must satisfy
the free-space Maxwell equation $\nabla \cdot {\bf E} = 0$.  If $f(r,z)$ is
not constant and $E_y = 0$, then we
must have nonzero $E_z$.  That is, the desired electric field has more 
than one vector component.

To deduce all components of the electric and magnetic fields of
a Gaussian laser beam from a single scalar wave function, we follow the 
suggestion of Davis \cite{Davis} and seek solutions for a vector potential 
{\bf A} that has only a single component.  We work in the Lorentz gauge
(and Gaussian units), so that the scalar potential $\Phi$ is related to the
vector potential by
\begin{equation}
\nabla \cdot {\bf A} + {1 \over c} {\partial \Phi \over \partial t} = 0.
\label{e1}
\end{equation}
The vector potential can therefore have a nonzero divergence, which permits
solutions having only a single component.  Of course, the electric and
magnetic fields can be deduced from the potentials via
\begin{equation}
{\bf E} = - \nabla \Phi - {1 \over c} {\partial {\bf A} \over \partial t},
\label{e2}
\end{equation} 
and
\begin{equation}
{\bf B} = \nabla \times {\bf A}.
\label{e2a}
\end{equation}
For this, the scalar potential must first be deduced from the vector potential
using the Lorentz condition (\ref{e1}).

The vector potential satisfies the free-space wave equation,
\begin{equation}
\nabla^2{\bf A} = {1 \over c^2} {\partial^2 {\bf A} \over \partial t^2}.
\label{e3}
\end{equation}
We seek a solution in which the vector potential is described by a single
component $A_j$ that propagates in the $+z$ direction with the form
\begin{equation}
A_j({\bf r},t) = \psi(r_\perp,z) g(\varphi) e^{i\varphi},
\label{e4}
\end{equation}
where the spatial envelope $\psi$ is azimuthally symmetric,
$r_\perp = \sqrt{x^2 + y^2}$, $g$ is the temporal pulse shape, and
the phase $\varphi$ is given by
\begin{equation}
\varphi = kz - \omega t.
\label{e5}
\end{equation}

Inserting trial solution (\ref{e4}) into the wave equation (\ref{e3})
we find that 
\begin{equation}
\nabla^2 \psi + 2ik{\partial\psi \over \partial z} \left( 1 - {i g' \over g}
\right) = 0,
\label{e7}
\end{equation}
where $g' = dg/d\varphi$.

\subsection{A Condition on the Temporal Pulse Shape $g(\varphi)$}

Since $\psi$ is a function of {\bf r} while $g$ and $g'$ are functions of
the phase $\varphi$, eq.~(\ref{e7}) cannot be satisfied in general.
Often the discussion is restricted to the case where $g' = 0$, \ie, to
continuous waves.  For a  pulsed laser beam, $g$ must obey 
\begin{equation}
\abs{g' \over g} \ll 1
\label{e8}
\end{equation}
for eq.~(\ref{e7}) to be consistent.

It is noteworthy that a ``Gaussian" laser beam cannot have a Gaussian 
temporal pulse.  That is, if $g = 
\exp[-(\varphi/\varphi_0)^2]$, then $\abs{g'/g} = 2 \abs{\varphi}/\varphi_0^2$,
which does not satisfy condition (\ref{e8}) for $|{\varphi}|$ large compared
to the characteristic pulsewidth $\varphi_0 = \omega \Delta t$, \ie, in the 
tails of the pulse.

A more appropriate form for a pulsed beam is a hyperbolic secant (as
arises in studies of solitons):
\begin{equation}
g(\varphi) = \hbox{sech} \left( {\varphi \over \varphi_0} \right).
\label{e9}
\end{equation}
Then,
$\abs{g'/g} = (1/\varphi_0) \abs{\tanh(\varphi/\varphi_0)}$, which is less 
than one everywhere provided that $\varphi_0 \gg 1$.

\subsection{The Paraxial Wave Equation}

In the remainder of this paper, we suppose that condition (\ref{e8}) is
satisfied.  Then, the differential equation (\ref{e7}) for the spatial
envelope function $\psi$ becomes
\begin{equation}
\nabla^2 \psi + 2ik{\partial\psi \over \partial z} = 0.
\label{e10}
\end{equation}

The function $\psi$ can and should be expressed in terms of three geometric
parameters of a focused beam, the diffraction angle $\theta_0$, the waist
$w_0$, and the depth of focus (Rayleigh range) $z_0$, which are related by
\begin{equation}
\theta_0 = {w_0 \over z_0} = {2 \over kw_0},
\qquad \mbox{and} \qquad
z_0 = {k w_0^2 \over 2} = {2 \over k \theta_0^2}.
\label{e11}
\end{equation}
We therefore work in the scaled coordinates
\begin{equation}
\xi = {x \over w_0}, \qquad 
\upsilon = {y \over w_0}, \qquad 
\rho^2 = {r_\perp^2 \over w_0^2} = \xi^2 + \upsilon^2, 
\qquad \hbox{and} \qquad 
\varsigma = {z \over z_0},
\label{e12}
\end{equation}
Changing variables and noting relations (\ref{e11}), eq.~(\ref{e10}) takes the
form
\begin{equation}
\nabla^2_\perp \psi + 4i {\partial\psi \over \partial\varsigma} + \theta_0^2 
{\partial^2\psi \over \partial\varsigma^2} = 0, 
\label{e13}
\end{equation}
where
\begin{equation}
\nabla^2_\perp \psi = 
{\partial^2 \psi \over \partial\xi^2} + 
{\partial^2 \psi \over \partial\upsilon^2}
= {1 \over \rho}{\partial \over \partial\rho} \left( \rho {\partial \psi
\over \partial \rho} \right),
\label{e14}
\end{equation}
since $\psi$ is independent of the azimuth $\phi$.

The form of eq.~(\ref{e13}) suggests the series expansion
\begin{equation}
\psi = \psi_0 + \theta_0^2 \psi_2 + \theta_0^4 \psi_4 + ...
\label{e15}
\end{equation}
in terms of the small parameter $\theta_0^2.$
Inserting this into eq.~(\ref{e13}) and collecting terms of order 
$\theta_0^0$ and $\theta_0^2$, we find
\begin{equation}
\nabla^2_\perp \psi_0 + 4i {\partial \psi_0 \over \partial\varsigma} = 0,
\label{e16}
\end{equation}
and
\begin{equation}
\nabla^2_\perp \psi_2 + 4i {\partial \psi_2 \over \partial\varsigma} = 
- {\partial^2 \psi_0 \over \partial\varsigma^2},
\label{e17}
\end{equation}
\etc\ \  

Equation (\ref{e16}) is called the the paraxial wave
equation, whose solution is well-known to be
\begin{equation}
\psi_0 = f e^{-f\rho^2},
\label{e18}
\end{equation}
where
\begin{equation}
f = {1 \over 1 + i \varsigma} = { 1 - i\varsigma \over 1 + \varsigma^2} = 
{e^{-i\tan^{-1} \varsigma} \over \sqrt{1 + \varsigma^2}}.
\label{e19}
\end{equation}
The factor $e^{-i\tan^{-1} \varsigma}$ in $f$ is the so-called Guoy phase shift
\cite{Siegman},
which changes from 0 to $\pi/2$ as $z$ varies from 0 to $\infty$, with the most
rapid change near the $z_0$.

The solution to eq.~(\ref{e17}) for $\psi_2$ has been given in \cite{Davis},
and that for $\psi_4$ has been discussed in \cite{Barton}.

With the lowest-order spatial function $\psi_0$ in hand, we are nearly ready to
display the electric and magnetic fields of the corresponding Gaussian beams.
But first, we need the scalar potential $\Phi$, which we suppose has the
form
\begin{equation}
\Phi({\bf r},t) = \Phi({\bf r}) g({\varphi}) e^{i\varphi},
\label{e20}
\end{equation}
similar to that of the vector potential.
Then,
\begin{equation}
{\partial \Phi \over \partial t}  = - \omega \Phi \left( 1 - {i g' \over g}
\right) \approx - \omega \Phi,
\label{e21}
\end{equation}
assuming condition (\ref{e8}) to be satisfied.  In that case,
\begin{equation}
\Phi = -{i \over k} \nabla \cdot {\bf A},
\label{e22}
\end{equation}
according to the Lorentz condition (\ref{e1}).  The electric field 
is then given by
\begin{equation}
{\bf E} = - {\bf \nabla} \Phi - {1 \over c} {\partial {\bf A} \over \partial t}
\approx  ik \left[ {\bf A} + {1 \over k^2} {\bf \nabla} ({\bf \nabla} \cdot 
{\bf A}) \right], 
\label{e23}
\end{equation}
in view of condition (\ref{e8}).  Note that $(1/k)\partial/\partial x = (\theta_0
/ 2) \partial / \partial \xi$, \etc, according to eqs.~(\ref{e11})-(\ref{e12}).

\subsection{Linearly Polarized Gaussian Beams}

\begin{figure}[htp]  
\begin{center}
\includegraphics[width=4in, angle=0, clip]{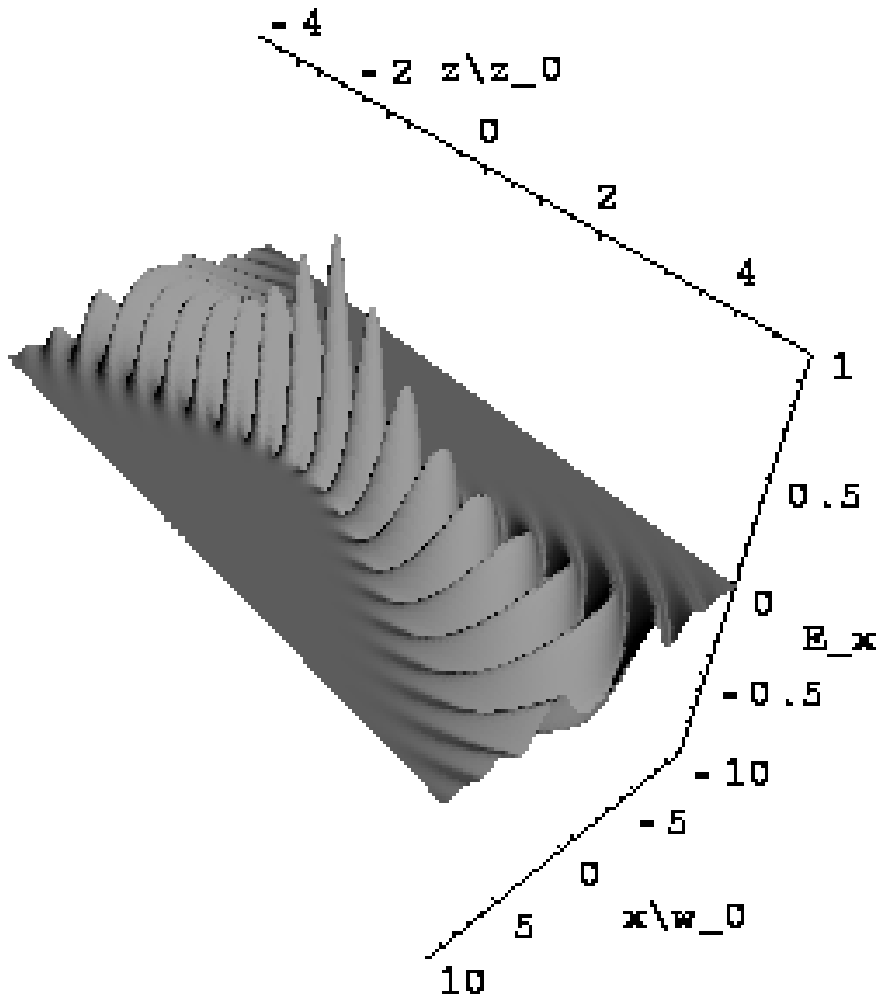}
\parbox{5.5in} 
{\caption[ Short caption for table of contents ]
{\label{exl} The electric field $E_x(x,0,z)$ of a linearly polarized Gaussian
beam with diffraction angle $\theta_0 = 0.45$, according to eq.~(\ref{e26a}).
}}
\end{center}
\end{figure}

\begin{figure}[htp]  
\begin{center}
\includegraphics[width=4in, angle=0, clip]{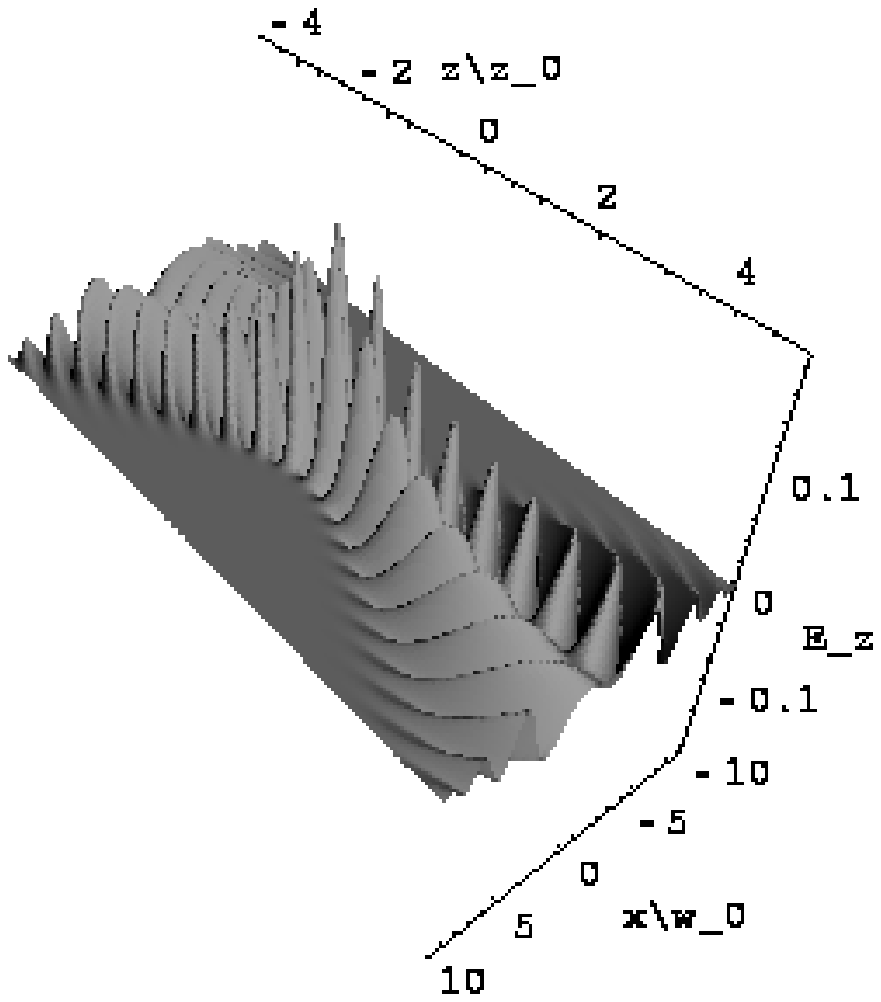}
\parbox{5.5in} 
{\caption[ Short caption for table of contents ]
{\label{ezl} The electric field $E_z(x,0,z)$ of a linearly polarized Gaussian
beam with diffraction angle $\theta_0 = 0.45$, according to eq.~(\ref{e26a}).
}}
\end{center}
\end{figure}

Taking the scalar wave function (\ref{e18}) to be the $x$ component
of the vector potential, 
\begin{equation}
A_x = {E_0 \over ik} \psi_0 g(\varphi) e^{i\varphi},
\qquad A_y = A_z = 0,
\label{e24}
\end{equation}
the corresponding electric and magnetic fields
are found from eqs.~(\ref{e2a}), (\ref{e23}) and (\ref{e24}) to be the 
familiar forms of a linearly polarized Gaussian beam,
\begin{eqnarray}
E_x & = & E_0 \psi_0 g e^{i\varphi} + {\cal O}(\theta_0^2)
\approx E_0 f e^{-f \rho^2} g e^{i \varphi}
\nonumber \\
& = & {E_0 e^{- \rho^2 / (1 + \varsigma^2)} g(\varphi) \over 
\sqrt{1 + \varsigma^2}}
e^{i [kz  + \varsigma \rho^2 / (1 + \varsigma^2) - \omega t - \tan^{-1}
\varsigma ]},
\nonumber \\
& = & {E_0 e^{- r_\perp^2 / w^2(z) } g(\varphi) \over 
\sqrt{1 + z^2/z_0^2}} e^{i \{kz[1 + r_\perp^2 / 2(z^2 + z_0^2)]  - \omega t 
- \tan^{-1} (z/z_0) \} },
\nonumber \\
E_y & = &  0, 
\label{e25} \\
E_z & = & {i\theta_0 E_0 \over 2} {\partial \psi_0 \over \partial \xi} g 
e^{i\varphi} + {\cal O}(\theta_0^3) \approx - i\theta_0 f\xi E_x,
\nonumber
\end{eqnarray}
\begin{eqnarray}
B_x & = &  0,
\nonumber \\
B_y & = & E_x,
\label{e26} \\
B_z & = & {i\theta_0 E_0 \over 2} {\partial \psi_0 \over \partial\upsilon} g
e^{i\varphi} 
= - i\theta_0 f\upsilon E_x,
\nonumber
\end{eqnarray}
where
\begin{equation}
w(z) = w_0 \sqrt{1 +z^2 / z_0^2}
\label{e26a}
\end{equation}
is the characteristic transverse size of the beam at position $z$.  
Near the focus ($r_\perp \lsim w_0, \abs{z} < z_0$), the beam is a plane wave,
\begin{equation}
E_x \approx E_0 e^{- r_\perp^2 / w_0^2} e^{i(kz - \omega t - z / z_0)}, 
\qquad E_z \approx  \theta_0 {x \over w_0} E_0 e^{- r_\perp^2 / w_0^2} 
 e^{i(kz - \omega t - 2 z / z_0 - \pi / 2)}, 
\label{e26b}
\end{equation}
For large $z$,
\begin{equation}
E_x \approx E_0 e^{- \theta^2 / \theta_0^2} {e^{i(k r - \omega t - \pi / 2)} 
\over r}, \qquad E_z \approx - {x \over r} E_x,
\label{e26c}
\end{equation}
where $r = \sqrt{r_\perp^2 + z^2}$ and $\theta \approx r_\perp / r$,
which describes a linearly polarized spherical wave of extent $\theta_0$ about
the $z$ axis.  The fields $E_x$ and $E_z$, \ie, the real parts of 
eqs.~(\ref{e26c}), are shown in Figs.~\ref{exl} and \ref{ezl}.

The fields (\ref{e25})-(\ref{e26})
 satisfy $\nabla \cdot {\bf E} = 0 = 
\nabla \cdot {\bf B}$ plus terms of order $\theta_0^2$.

Clearly, a vector potential with only a $y$ component of form similar to
eq.~(\ref{e24})
leads to the lowest-order Gaussian beam with linear polarization in the $y$
direction.

\subsection{The Lowest-Order Axicon Beam}

\begin{figure}[htp]  
\begin{center}
\includegraphics[width=4in, angle=0, clip]{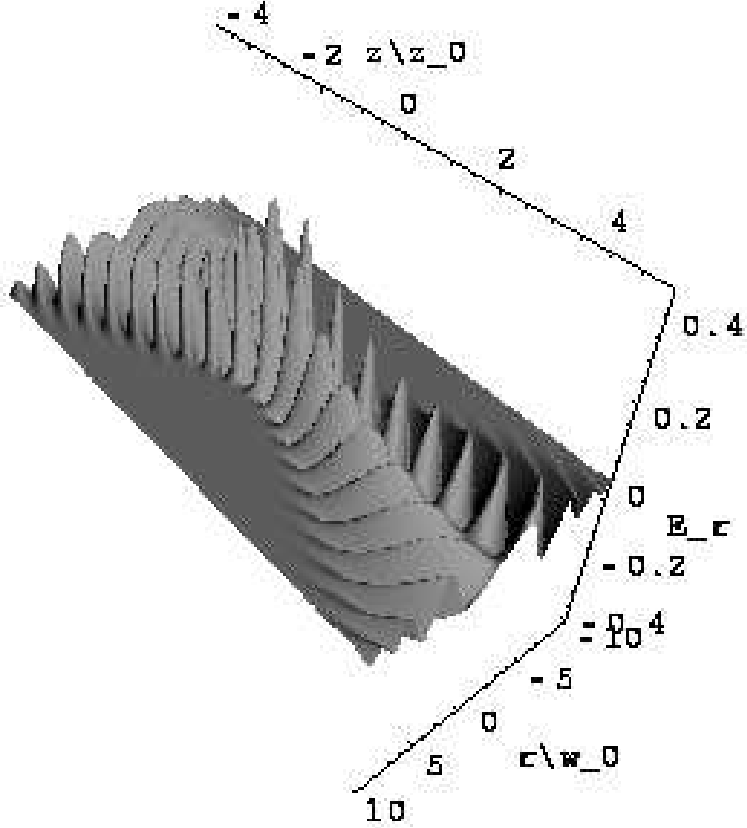}
\parbox{5.5in} 
{\caption[ Short caption for table of contents ]
{\label{era} The electric field $E_r(r_\perp,0,z)$ of an axicon Gaussian
beam with diffraction angle $\theta_0 = 0.45$, according to eq.~(\ref{e29}).
}}
\end{center}
\end{figure}

\begin{figure}[htp]  
\begin{center}
\includegraphics[width=4in, angle=0, clip]{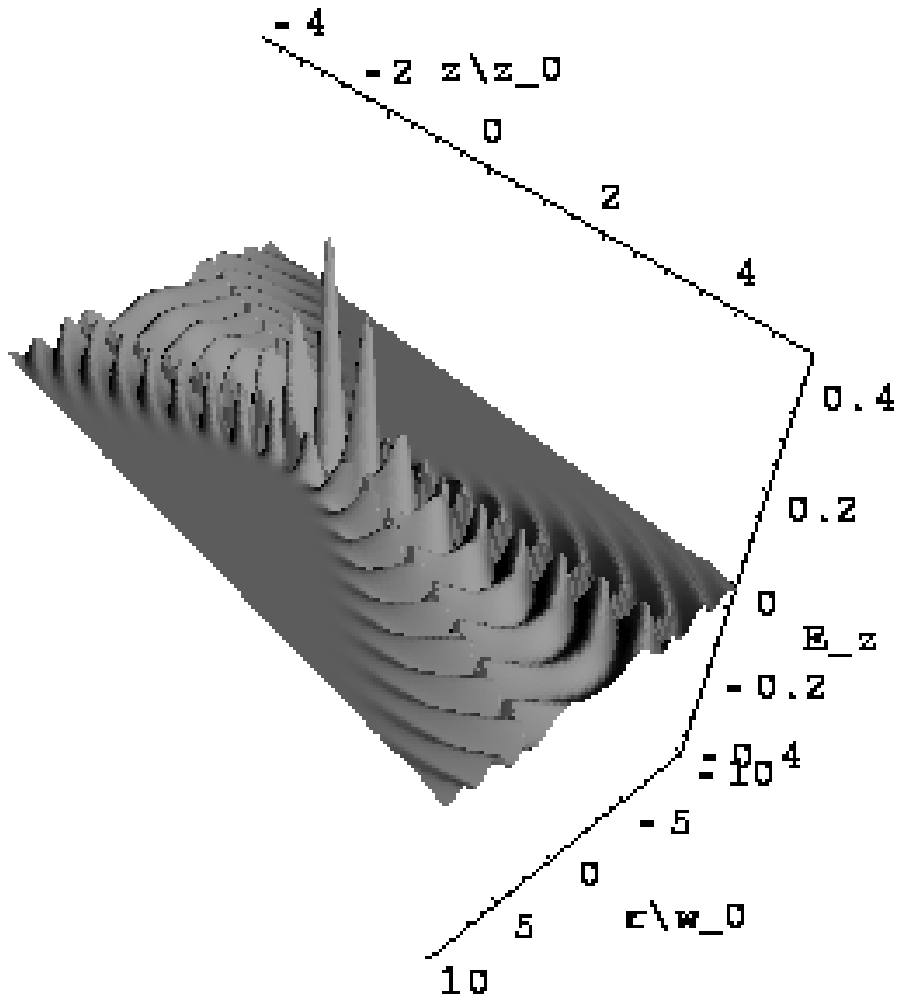}
\parbox{5.5in} 
{\caption[ Short caption for table of contents ]
{\label{eza} The electric field $E_z(r_\perp,0,z)$ of an axicon Gaussian
beam with diffraction angle $\theta_0 = 0.45$, according to eq.~(\ref{e29}).
}}
\end{center}
\end{figure}

An advantage of our solution based on the vector potential is that we also can
consider the case that only $A_z$ is nonzero and has the form (\ref{e18}),
\begin{equation}
A_x = A_y = 0, \qquad A_z 
= {E_0 \over k \theta_0} f e^{-f \rho^2} g e^{i(k z - \omega t)}.
\label{e27}
\end{equation}
Then,
\begin{equation}
\nabla \cdot {\bf A} = {\partial A_z \over \partial z} 
\approx i k A_z \left[ 1 -  {\theta_0^2 \over 2 } f (1 - f \rho^2) \right],
\label{e28}
\end{equation}
using eqs.~(\ref{e11})-(\ref{e12}) and the fact that $df/d \varsigma =
- i f^2$, which follows from eq.~(\ref{e19}).
Anticipating that the electric field has radial polarization, we work in 
cylindrical coordinates, $(r_\perp,\phi,z)$, and find  from eqs.~(\ref{e2a}), 
(\ref{e23}), (\ref{e27}) and (\ref{e28}) that
\begin{eqnarray}
E_\perp & = & E_0 \rho f^2 e^{-f \rho^2} g e^{i \varphi} + {\cal O}(\theta_0^2),
\nonumber \\
E_\phi & = & 0,
\label{e29} \\
E_z & = & i \theta_0 E_0 f^2 (1 - f \rho^2) e^{-f \rho^2} g e^{i \varphi} + 
{\cal O}(\theta_0^3).
\nonumber
\end{eqnarray}
The magnetic field is
\begin{equation}
B_\perp = 0, \qquad B_\phi = E_\perp, \qquad B_z = 0.
\label{e30}
\end{equation}
The fields $E_x$ and $E_z$ are shown in Figs.~\ref{era} and
\ref{eza}.  The dislocation seen in Fig.~\ref{eza} for 
$\rho \approx \varsigma$ is due to the factor $1 - f \rho^2$ that arises in
the paraxial 
approximation, and would, I believe, be smoothed out on keeping higher-order
terms in the expansion (\ref{e15}).

The transverse electric field is radially polarized and vanishes on the axis.
The longitudinal electric field is nonzero on the axis.
Near the focus, $E_z \approx i \theta_0 E_0$ and the peak radial field is
$E_0/\sqrt{2e} = 0.42 E_0$.
For large $z$, $E_\perp$ peaks at $\rho = \varsigma/\sqrt{2}$,
corresponding to polar angle $\theta = \theta_0/\sqrt{2}$.  For angles near 
this,
$\abs{E_\perp} \approx \rho \abs{f}^2 \approx 1/z$, as expected
in the far zone.  In this region, the ratio of the longitudinal to transverse
fields is $E_z / E_\perp \approx -i \theta_0 f \rho \approx - r_\perp / z$, as
expected for a spherical wave front.

The factor $f^2$ in the fields implies a Guoy phase shift of $e^{-2 i
\tan^{-1} \varsigma}$, which is twice that of the lowest-order linearly 
polarized beams.

It is noteworthy that the simplest axicon mode (\ref{e29})-(\ref{e30})
is not a member of the set of Gaussian modes based on Laguerre polynomials
in cylindrical coordinates (see, for example, sec.~3.3b of \cite {Kogelnik}).

\subsection{Guided Axicon Beam}

We could also consider the vector potential
\begin{equation}
A_{r_\perp} \propto \psi_0 g e^{i\varphi}, \qquad A_\phi = A_z = 0,
\label{e31}
\end{equation}
which leads to the electric and magnetic fields
\begin{equation}
E_r = E_0 f e^{- f \rho^2} g e^{i \varphi},\ E_\phi = 0,\ 
E_z = - i \theta_0 f \rho E_r,
\qquad
B_r = 0,\ B_\phi = E_r,\ B_z = 0,
\label{e33}
\end{equation}
and the potential
\begin{equation}
A_{r_\perp} = 0, \qquad A_\phi \propto \psi_0 g e^{i\varphi}, \qquad  A_z = 0,
\label{e32}
\end{equation}
which leads to
\begin{equation}
E_r = 0,\ E_\phi = E_0 f e^{- f \rho^2} g e^{i \varphi},\ E_z = 0,
\qquad
B_r = - E_\phi,\ B_\phi = 0,\
 B_z = - i \theta_0 {1 - 2 f \rho^2 \over 2 \rho}  E_\phi.
\label{e34}
\end{equation}
The case of eqs.~(\ref{e32})-(\ref{e34}) is unphysical
due to the blowup of $B_z$ as $r_\perp \to 0$.

The fields of eqs.~(\ref{e31})-(\ref{e33}) do not satisfy $\nabla \cdot
{\bf E} = 0$ at $r_\perp = 0$, and so cannot correspond to a free-space wave.
However, these fields could be supported by a wire, and represent a TM
axicon guided
cylindrical wave with a focal point.  This is in contrast to guided plane waves 
whose radial
profile is independent of $z$ \cite{Sommerfeld,Stratton}.  Guided axicon
beams might
find application when a focused beam is desired at a point
where a system of lenses and mirrors cannot conveniently deliver the
optical axis, or in wire-guided atomic traps \cite{Denschlag}.
Figures \ref{exl} and \ref{ezl} show
the functional form of the guided axicon beam (\ref{e33}), 
when coordinate $x$ is reinterpreted as $r_\perp$.


\end{document}